\documentclass[epj]{mysvjour}
\usepackage{graphicx,epstopdf,multirow,amsmath,rotating,subfigure,isotope,gensymb,amssymb}
\usepackage{graphicx}
\usepackage{dcolumn}
\usepackage{bm}
\usepackage{tabularx}
\usepackage{makecell}
\usepackage[export]{adjustbox}
\usepackage{float}
\usepackage{braket}
\usepackage{lscape}
\usepackage{lipsum}
\usepackage{longtable}

\usepackage[T1]{fontenc}
\usepackage[utf8]{inputenc} 

\usepackage{latexsym}
\usepackage{graphics}
\usepackage{xcolor}   

\usepackage{hyperref}
\hypersetup{
    colorlinks=true, 
    linktoc=all,     
    urlcolor  = blue,
    citecolor = blue,
    linkcolor=blue,  
}
\makeatletter
\newcommand{\colorcaption}[2][]{%
  \begingroup%
  \renewcommand{\@caption@fignum@sep}{ (Color online). }%
  \caption[#1]{#2}%
  \endgroup%
}
\makeatother

\usepackage{amsmath}
\newcommand{\orcid}[1]{\href{https://orcid.org/#1}{\hskip2pt\includegraphics[width=9pt]{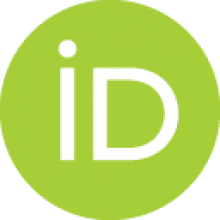}}}
\begin{document}
\title{Large-scale shell-model investigation of $2\nu$ECEC in $^{132}$Ba and $^{78}$Kr}
\author{Deepak Patel$^1$\orcid{0000-0002-7669-1907} \thanks{d{\textunderscore}patel@ph.iitr.ac.in}, Praveen C. Srivastava$^1$\orcid{0000-0001-8719-1548} \thanks{praveen.srivastava@ph.iitr.ac.in}}
\institute{$^{1}$Department of Physics, Indian Institute of Technology Roorkee, Roorkee 247 667, 
INDIA }

\date{\today}

\abstract{We present a theoretical investigation of two-neutrino double electron capture ($2\nu$ECEC) in $^{132}$Ba and $^{78}$Kr based on large-scale shell-model calculations. The nuclear matrix elements (NMEs) for the $2\nu$ECEC process in $^{132}$Ba and $^{78}$Kr are calculated using the SN100PN and GWBXG effective interactions, respectively. The reliability of the employed interactions is first examined through a comparison of the calculated and experimental spectroscopic properties of the parent, intermediate, and granddaughter nuclei involved in the decay. We also examine the cumulative contribution of the $2\nu$ECEC NME with respect to the $1^+$ state energies in the intermediate nuclei. The present results provide an updated and improved shell-model estimate of the $2\nu$ECEC NME and half-life for $^{78}$Kr relative to earlier studies, and a baseline theoretical prediction for $^{132}$Ba that may assist future experimental efforts in constraining this rare decay mode.}

\authorrunning{D. Patel and P.C. Srivastava}
\maketitle

\section{Introduction} \label{section-A}

The study of second-order weak interaction processes in atomic nuclei, such as double-beta ($\beta\beta$) decay and double electron capture (ECEC), provides a powerful tool for probing the fundamental properties of neutrinos and weak interactions, as well as for gaining insight into nuclear structure and testing the validity of conservation laws \cite{Brown,Vogel,Aprile2,Abe,Suhonen,Hinohara,Rodriguez1,Poves,Engel,Menendez,Coello}. These rare nuclear decays occur in two distinct modes: the two-neutrino ($2\nu$) mode, which conserves lepton number and is allowed within the framework of standard model \cite{Deppisch}, and the neutrinoless ($0\nu$) mode, which violates lepton number conservation and, if observed, would have profound implications for our understanding of the nature of neutrinos \cite{Cirigliano,Rodriguez,Shimizu1,Iwata,Coraggio}. Although a controversial claim of evidence for $0\nu\beta\beta$ decay in $^{76}$Ge was reported by part of the Heidelberg-Moscow collaboration \cite{Klapdor}, this result was later refuted by the more sensitive GERDA \cite{Agostini} and Majorana Demonstrator \cite{Arnquist} experiments. The $2\nu$ mode, which is experimentally established and theoretically well-understood, serves as a vital benchmark for testing nuclear structure models, provides important background information for $0\nu\beta\beta$ and $0\nu$ECEC searches, and enables probing of the effective value of the weak axial-vector coupling constant \cite{Ackerman,Adams,Jokiniemi,Patel_76Ge,Kostensalo1}. In this work, we focus on the $2\nu$ECEC process, which, although allowed by standard model, has been far less explored compared to its $2\nu\beta\beta$ counterpart, primarily due to its typically longer half-lives and lower $Q$-values \cite{Meshik}. It also plays a crucial role in benchmarking nuclear matrix element (NME) calculations that are also relevant for neutrinoless modes.

Among the nuclei that are energetically allowed to undergo $2\nu$ECEC, $^{132}$Ba is of particular interest. It is one of the few nuclei for which a geochemical signature of the $2\nu$ECEC process has been reported, although only a lower limit on the half-life has been established to date \cite{Meshik}. Experimental investigations of $2\nu$ECEC have also been pursued in a limited number of other potential candidates. In particular, the process has been geochemically confirmed in $^{130}$Ba \cite{Meshik,Pujol}, and direct observations have recently been achieved for $^{78}$Kr using low-background gas proportional counters \cite{Gavrilyuk,Ratkevich} and for $^{124}$Xe in large-scale liquid-xenon detectors with XMASS \cite{Abe}, XENON \cite{Aprile,Aprile1,Aprile2}, the PandaX-4T \cite{Bo}, and the LZ \cite{Aalbers} Collaborations. The rapid development of ultra-low-background detection techniques and rare-event search experiments has therefore renewed interest in exploring additional candidate nuclei for $2\nu$ECEC. In this context, reliable theoretical predictions of the NMEs and half-lives for $^{132}$Ba are important to guide future experimental searches, estimate the required sensitivity, and assess the feasibility of detecting this process.

Existing theoretical estimates of the $2\nu$ECEC half-life exhibit significant model dependence, particularly in the treatment of NMEs \cite{Suhonen,Suhonen5}. This underscores the importance of using robust and independently validated methods. In this context, the nuclear shell model (NSM) has shown considerable success in describing spectroscopic properties \cite{Brown1} and double-beta decay observables in the $A \sim 130$ mass region, including in nuclei such as $^{124}$Xe, $^{128,130}$Te, and $^{136}$Xe \cite{Coello,Horoi1,Caurier,Patel_NPA}. In particular, in our previous work \cite{Patel_NPA}, the $2\nu\beta\beta$ decay of $^{124}$Sn, $^{128}$Te, $^{130}$Te, and $^{136}$Xe was studied within the NSM using the same SN100PN interaction \cite{Brown1} employed in the present work for $^{132}$Ba. In that work \cite{Patel_NPA}, we excluded the $h_{11/2}$ orbital from the proton model space for the calculation of $^{130}$Te. Using the computed NME (0.0516) and an effective axial-vector coupling in the range $g_{\rm A}^{\rm eff} = 0.94$-$0.72$, the resulting half-life lies in the range $(3.37$-$9.78)\times10^{20}$ yr, in good agreement with the experimental value of $(7.91 \pm 0.21)\times10^{20}$ yr \cite{Barabash}. This provides confidence to use present effective interaction to investigate the $2\nu$ECEC process in $^{132}$Ba. Although other nuclear models, such as the quasiparticle random-phase approximation (QRPA) \cite{Suhonen5,Yousef}, interacting boson model (IBM) \cite{Yoshida,Nomura}, Hartree-Fock-Bogoliubov (HFB) model \cite{Raina}, and effective field theory \cite{Coello,Coello1,Jokiniemi1}, have also been used previously to study the $2\nu\beta\beta$ decay and $2\nu$ECEC processes.

In this work, we carry out a detailed investigation of the $2\nu$ECEC process in $^{132}$Ba using large-scale shell-model calculations. To ensure a more precise estimation of the NME, we first aim to validate the reliability of the employed effective interaction by examining its predictive power for spectroscopic properties compared to the experimental data. Using a large model space and effective interaction tailored for this mass region, we compute the NME and half-life for the $2\nu$ECEC process in $^{132}$Ba. To the best of our knowledge, this work presents the first shell-model calculation of the NME for the $2\nu$ECEC process in $^{132}$Ba. In our earlier work, we investigated the $2\nu$ECEC process in $^{78}$Kr using a shell-model configuration limited to the proton ($Z$) valence space $0f_{5/2}1p0g_{9/2}$ and the neutron ($N$) space $1p_{1/2}0g_{9/2}$ \cite{Patel_78Kr}. Thus, in the present study, we also aim to provide a more accurate estimation of the NME for $2\nu$ECEC process in $^{78}$Kr by extending the neutron model space, which includes the excitations into the $0g_{7/2}1d2s$ orbitals above the $N=50$ shell closure.

This paper is structured as follows: Section \ref{section2} offers a concise overview of the theoretical framework employed in our calculations. Section \ref{section3} presents the results of the shell-model analyses, including discussions on the energy spectra, transition probabilities, NMEs, and the extracted half-lives. Finally, Section \ref{section4} summarizes our findings and provides the conclusions.

\begin{figure}
	\centering
	\includegraphics[width=83mm]{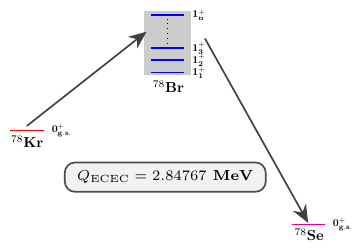}
	\includegraphics[width=83mm]{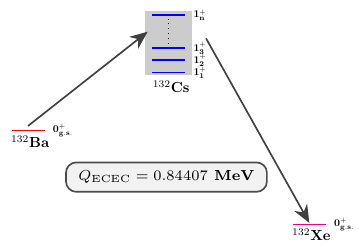}
	\caption{\label{scheme_decay} Nuclear level and decay scheme for $2\nu$ECEC process in $^{78}$Kr (upper panel) and $^{132}$Ba (lower panel).}
\end{figure}

\section{Formalism} \label{section2}

The half-life for the $2\nu$ECEC process can be expressed as
\begin{equation}
	\begin{aligned}
		(T_{1/2}^{2\nu})^{-1} &= G_{2\nu}^{\rm ECEC}(g_{\rm A}^{\rm eff})^4 \left|M_{2\nu}^{\rm GT}\right|^2.
	\end{aligned}
	\label{GT+Fermi:NME}
\end{equation}

Here, $G_{2\nu}^{\rm ECEC}$ denotes the phase-space factor \cite{Nitescu,Kotila}, while $g_{\rm A}^{\rm eff}$ represents the effective axial-vector coupling constant \cite{Suhonen4}. The NME for the 2$\nu$ECEC process corresponding to the Gamow-Teller (GT) transition is defined as follows \cite{Stefanik,Nitescu1}
\begin{equation}
	M_{2\nu}^{\rm GT}=m_e \sum_{n}\frac{\langle 0_{\rm g.s.}^{(f)}||\sum_{a}\sigma_a\tau^{+}_a||1_n^+\rangle \langle 1_n^+||\sum_{b}\sigma_b\tau^{+}_b||0_{\rm g.s.}^{(i)}\rangle}{E(1^+_n)-(E_i+E_f)/2}.
	\label{eq2}
\end{equation}

In Eq. (\ref{eq2}), $E(1^+_n)$ denotes the energy of $n^{\rm th}$ $1^+$ state in the intermediate nucleus. $E_i$ and $E_f$ stand for the ground state (g.s.) energies of the parent and granddaughter nuclei, respectively. The symbol $m_e$ represents the rest mass of the electron. The notations $0_{\rm g.s.}^{(i)}$ and $0_{\rm g.s.}^{(f)}$ correspond to the ground states of the parent and granddaughter nuclei, respectively. The Pauli spin matrices are denoted by $\sigma_a$ and $\sigma_b$, while $\tau^+_a$ and $\tau^+_b$ stand for the isospin-raising operators. The reduced GT matrix elements are written as $\langle 0_{\rm g.s.}^{(f)} || \sum_{a} \sigma_a \tau^+_a || 1_n^+ \rangle$ and $\langle 1_n^+ || \sum_{b} \sigma_b \tau^+_b || 0_{\rm g.s.}^{(i)} \rangle$. It is important to emphasize that, within the shell-model framework, the $g_{\rm A}^{\rm eff}$ appearing in Eq. (\ref{GT+Fermi:NME}) parametrizes the phenomenological renormalization of the spin-isospin operator $\sum_{a} \sigma_a \tau^+_a$ entering the NME of Eq. (\ref{eq2}). In practical applications, the axial-vector coupling is factorized in the half-life expression, while the NME $M_{2\nu}^{\rm GT}$ in Eq. (\ref{eq2}) is evaluated using the standard one-body GT operator. The nuclear level and decay schemes for the $2\nu$ECEC processes in $^{78}$Kr and $^{132}$Ba are shown in Fig. \ref{scheme_decay}.

It is well known that weak nuclear transitions receive contributions from two-body (meson-exchange) currents and correlations beyond the valence space, which lead to a suppression of spin-isospin matrix elements \cite{Menendez,Ekstrom,Zhou}. In particular, recent \textit {ab initio} calculations have demonstrated that two-body currents, together with strong nuclear correlations, produce a substantial reduction of GT matrix elements and observed $\beta$-decay strengths can be largely reproduced without additional quenching \cite{Gysbers}. Therefore, neglecting these contributions may result in an overestimation of the calculated $2\nu$ECEC NME. Accordingly, the use of an effective axial coupling, $g_{\rm A}^{\rm eff}$, provides a phenomenological way to incorporate these missing many-body contributions. Consequently, although two-body currents are not explicitly included, their impact is partially absorbed into the adopted $g_{\rm A}^{\rm eff}$, and the resulting half-life predictions are expected to remain reasonable.

\begin{figure*}
	\centering
	\includegraphics[width=40mm]{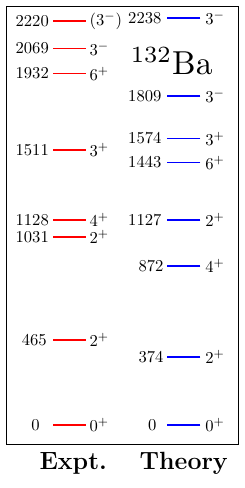}
	\includegraphics[width=40mm]{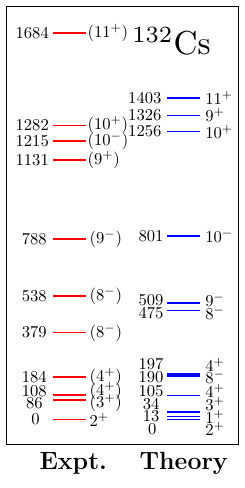}
	\includegraphics[width=40mm]{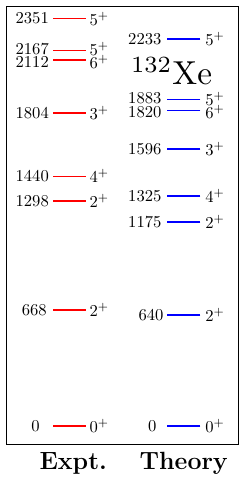}
	\caption{\label{spectrum} Comparison between the theoretical and experimental \cite{NNDC} energy levels (in keV) in the $^{132}$Ba, $^{132}$Cs and $^{132}$Xe.}
\end{figure*}

In our study of the $2\nu$ECEC process in $^{132}$Ba, we employed the SN100PN effective interaction \cite{Brown1}, which includes the $0g_{7/2}1d2s0h_{11/2}$ orbitals for both protons and neutrons within the 50-82 model space, taking $^{100}$Sn as an inert core. This interaction is based on a $G$-matrix derived from the CD-Bonn nucleon-nucleon potential \cite{Machleidt1}. Due to the large dimensions of the full model space, we introduced truncations to ensure computational feasibility. Specifically, the proton and neutron configurations are restricted to $\pi[(g_{7/2}d_{5/2})^{0-14}(d_{3/2}s_{1/2}h_{11/2})^{0-2}]$ and $\nu[(g_{7/2})^{8-8}(d_{5/2}d_{3/2}s_{1/2}h_{11/2})^{0-24}]$, respectively. We have included the $1^+$ states of the intermediate nucleus up to the excitation energies at which the cumulative NMEs reach the saturation level (5000 eigenstates).

For the calculation of the NME corresponding to the $2\nu$ECEC in $^{78}$Kr, we employed the same effective interaction (GWBXG) \cite{Hosaka,Ji,Gloeckner,Serduke} as used in our previous study on $2\nu$ECEC in $^{78}$Kr \cite{Patel_78Kr}. As discussed in Sec. \ref{section-A}, those earlier calculations were carried out without incorporating neutron orbitals above the $N=50$ shell closure. In the present work, we have extended the model space by allowing one-particle-one-hole ($1p$-$1h$) excitations across $N=50$, into the $0g_{7/2}$, $1d_{5/2}$, $1d_{3/2}$, and $2s_{1/2}$ neutron orbitals. The one-body transition densities (OBTDs) were computed with the NUSHELLX code \cite{Nushellx}. The KSHELL code \cite{KShell} was also utilized to calculate energy spectra and transition probabilities.

In the present study, the proton and neutron orbitals included in the model spaces for both $2\nu$ECEC candidates are not identical (for $^{132}$Ba, the $\nu g_{7/2}$ orbital is completely filled). Therefore, instead of the conventional Ikeda sum rule (ISR), we verify the effective Ikeda sum rule (EISR),	$3(N_{\rm core}-Z_{\rm core})+\sum B(\mathrm{GT}^-)-\sum B(\mathrm{GT}^+)=3(N-Z)$, where $N_{\rm core}$ and $Z_{\rm core}$ denote the neutron and proton numbers of the inert core. This relation follows from $3(N_{\rm act}-Z_{\rm act})=\sum B(\mathrm{GT}^-)-\sum B(\mathrm{GT}^+)$ discussed in Sec. 17.4.3 (Eq. 17.83) of Ref. \cite{Suh2007}, and was also adopted in our previous work \cite{Patel_76Ge}. For $^{132}$Ba and $^{132}$Xe, the calculated EISR values are 50.43 and 58.68, compared with the expected values 60 and 72, respectively; the deviations mainly arise from the truncation allowing only two protons in the $d_{3/2}sh_{11/2}$ orbitals. Similarly, for $^{78}$Se and $^{78}$Kr, the calculated EISR values are 33.52 and 25.88, compared with the expected values 30 and 18, respectively, where the latter shows some deviation from the expected value.

\section{Results and discussion} \label{section3}

In Sec. \ref{Sec-IIIA}, we assess the efficiency of the employed effective interaction in studying the $2\nu$ECEC process in $^{132}$Ba by comparing calculated spectroscopic properties of the parent, intermediate, and granddaughter nuclei, such as low-lying energy spectra and reduced quadrupole transition probabilities $B(E2)$, with the available experimental data. In Sec. \ref{Sec-IIIB}, we investigate the behavior of the cumulative GT NMEs with respect to the $1^+$ states of the intermediate nucleus $^{132}$Cs, as well as the corresponding half-life of $^{132}$Ba for the $2\nu$ECEC process. In Sec. \ref{Sec-IIIC}, we present updated results for the spectroscopic properties and the $2\nu$ECEC NME of $^{78}$Kr. The $Q_{\rm ECEC}$ (total energy released in the process) used in the NME calculations is taken from the Ref. \cite{Nitescu}.

\subsection{Analysis of the spectroscopic properties of $^{132}$Ba, $^{132}$Cs and $^{132}$Xe} \label{Sec-IIIA}

In Fig. \ref{spectrum}, we depict the comparison between theoretical and experimental \cite{NNDC} low-lying states of $^{132}$Ba, $^{132}$Cs and $^{132}$Xe. We found remarkable agreement between theoretical and experimental levels of the parent ($^{132}$Ba), intermediate ($^{132}$Cs), and granddaughter ($^{132}$Xe) nuclei of the $2\nu$ECEC process. The $0^+_1$, $2^+_1$, $4^+_1$, and $6^+_1$ states of $^{132}$Ba, which belong to the g.s. band, are quite fragmented in the shell-model calculations and are primarily dominated by two major configurations, $[\pi(d_{5/2}^2g_{7/2}^4) \otimes \nu(d_{5/2}^6d_{3/2}^2h_{11/2}^{10})]$ and $[\pi(d_{5/2}^2g_{7/2}^4) \otimes \nu(d_{5/2}^6d_{3/2}^2s_{1/2}^2h_{11/2}^{8})]$, each contributing $\lesssim$10\%. This fragmentation in the wavefunctions of these states leads to increased collectivity and enhanced $E2$ transitions between two consecutive states of the g.s. band, as predicted by theory and supported by the available experimental $B(E2)$ value reported in Table \ref{table:BE2}. The predicted spectroscopic quadrupole moments ($Q_s$) for the $2^+_1$, $4^+_1$, and $6^+_1$ states in $^{132}$Ba are negative. Using the relations $Q_s=[3K^2-J(J+1)]Q_0/[(J+1)(2J+3)]$ and $\beta_2=\sqrt{5\pi}Q_0/(3ZR^2)$, with $K=0$ for the $0^+_{\rm g.s.}$ band, also verified from the shell-model predicted $B(E2)$ ratios between considered consecutive states of the rotational band \cite{Patel_odd_Cd}, we obtain deformation parameter $\beta_2=+0.18$, $+0.19$, and $+0.18$ for the $2^+_1$, $4^+_1$, and $6^+_1$ states with $(e_\pi, e_\nu) = (1.6, 0.8)e$, respectively. Here, $J$ denotes the spin, $Q_0$ the intrinsic quadrupole moment, $Z$ the atomic number, and $R=1.2\times A^{1/3}$ fm. These results indicate a moderately prolate deformation, slightly larger than the previously estimated value ($\beta_2=+0.13$) \cite{Burnett}. In the case of $^{132}$Xe, the $0^+_1$, $2^+_1$, and $4^+_1$ states are characterized by the major configuration $[\pi(g_{7/2}^4) \otimes \nu(d_{5/2}^6d_{3/2}^2s_{1/2}^2h_{11/2}^{10})]$, contributing $\sim19\%$. These states exhibit strong $E2$ transitions as suggested by both the theoretical predictions and experimental data (see Table \ref{table:BE2}). In contrast, the $6^+_1$ state originates from the largest fraction $[\pi(g_{7/2}^3d_{5/2}^1) \otimes \nu(d_{5/2}^6d_{3/2}^2s_{1/2}^2h_{11/2}^{10})]$, with a contribution of 19.6\%. This structural change may lead to a slight reduction in the $B(E2)$ strength from the shell-model for the $6^+_1 \to 4^+_1$ transition compared to the $4^+_1 \to 2^+_1$ and $2^+_1 \to 0^+_1$ transitions. However, no $E2$ transition was observed between the $6^+_1$ and $4^+_1$ states in the previous experimental investigation \cite{Vogt}. Thus, we have calculated the $\beta_2$ values for only the $2^+_1$ and $4^+_1$ states, which are 0.02 and 0.03, respectively, indicating very small prolate deformation or a nearly spherical structure. The theoretical and experimental $E(4^+_1)/E(2^+_1)$ ratios also support the inferred deformations in $^{132}$Ba and $^{132}$Xe. For $^{132}_{55}$Cs$_{77}$, the shell-model calculations successfully reproduce the experimental ground and several excited states. Theoretical predictions for nuclei with both odd-$Z$ and odd-$N$ remain challenging, particularly in the medium- to heavy-mass regions. Nevertheless, the calculated energy levels for $^{132}$Cs show reasonably good agreement with the experimental data. At present, no experimental information on $B(E2)$ values in $^{132}$Cs is available in the NNDC database \cite{NNDC,NNDC_NUDAT}; therefore, only the shell-model predictions are presented in Table \ref{table:BE2}.

\begin{table} 
	\centering
	\caption{Comparison between theoretical and experimental \cite{NNDC_NUDAT} $B(E2)$ strengths [in Weisskopf units (W.u.)] for $^{132}$Ba, $^{132}$Cs and $^{132}$Xe. Theoretical $B(E2)$ values are calculated using two sets of effective charges $(e_\pi, e_\nu) = (1.6, 0.8)e$ \cite{Patel_odd_Cd,Honma} and $(e_\pi, e_\nu) = (1.7, 1.1)e$ \cite{Boelaert}. The $\hbar\omega=45A^{-\frac{1}{3}}-25A^{-\frac{2}{3}}$ MeV \cite{Suh2007} is used in the oscillator length $b = 197.33/\sqrt{940\times \hbar\omega[\rm MeV]}$ fm for the calculation of $B(E2)$ values. }
	\begin{tabular}{ccccc}
		\hline \hline
		Nucleus & $J^{\pi}_i \to J^{\pi}_f$ & \multicolumn{2}{c}{Theory} & Expt. \\
		\hline 
		
		&  & $(1.6, 0.8)e$ & $(1.7, 1.1)e$ &  \\
		
		\cline{3-4}
		
		$^{132}$Ba   & $2^+_1 \to 0^+_1$   & 38.3 & 52.7 & 43(4)   \\
		& $2^+_2 \to 0^+_1$ & 1.1 & 1.6 & 3.9(4) \\
		
		& $4^+_1 \to 2^+_1$ & 56.7 & 76.9 & - \\
		
		& $6^+_1 \to 4^+_1$ & 65.1 & 87.9 & - \\
		\hline \\
		
		$^{132}$Cs   & $3^+_1 \to 2^+_1$  & 7.9 & 11.0 & -   \\
		
		& $3^+_1 \to 1^+_1$  & 21.0 & 28.9 & -   \\
		
		& $4^+_1 \to 2^+_1$  & 14.3 & 20.2 & -   \\
		
		& $9^-_1 \to 8^-_1$  & 23.5 & 32.5 & -   \\
		
		& $10^-_1 \to 8^-_1$  & 23.0 & 32.3 & -   \\
		
		\hline \\
		
		$^{132}$Xe   & $2^+_1 \to 0^+_1$   & 19.9 & 28.2 & 23.1(15)   \\
		& $2^+_2 \to 0^+_1$   & 0.1 & 0.2 & 0.079(11)   \\
		& $2^+_2 \to 2^+_1$   & 26.4 & 38.3 & 41(4)   \\
		& $4^+_1 \to 2^+_1$   & 28.2 & 39.7 & 28.6(23)   \\
		& $6^+_1 \to 4^+_1$   & 15.3 & 21.2 & -   \\
		\hline \hline
	\end{tabular}
	\label{table:BE2}
\end{table}

The above discussion suggests that the employed effective interaction (SN100PN) successfully reproduces the spectroscopic properties of $^{132}$Ba, $^{132}$Cs, and $^{132}$Xe. Although the reproduction of low-lying excitation spectra constitutes a necessary validation of the shell-model wave functions but does not, by itself, ensure the accuracy of $2\nu$ECEC NME. As suggested in Ref. \cite{Coello}, even effective interactions that successfully describe spectroscopic properties require additional renormalization of the GT operator when applied to these decay processes, reflecting missing correlations beyond the valence space and higher-order many-body effects. Accordingly, the present spectroscopic agreement supports the reliability of the underlying nuclear wave functions, which form a reliable basis for the subsequent evaluation of the $2\nu$ECEC NME, while the limitations associated with the decay operator are treated through effective axial couplings in the half-life calculations.

\subsection{Cumulative NMEs and half-life of $^{132}$Ba} \label{Sec-IIIB}

\begin{figure}
	\centering
	\includegraphics[width=86mm]{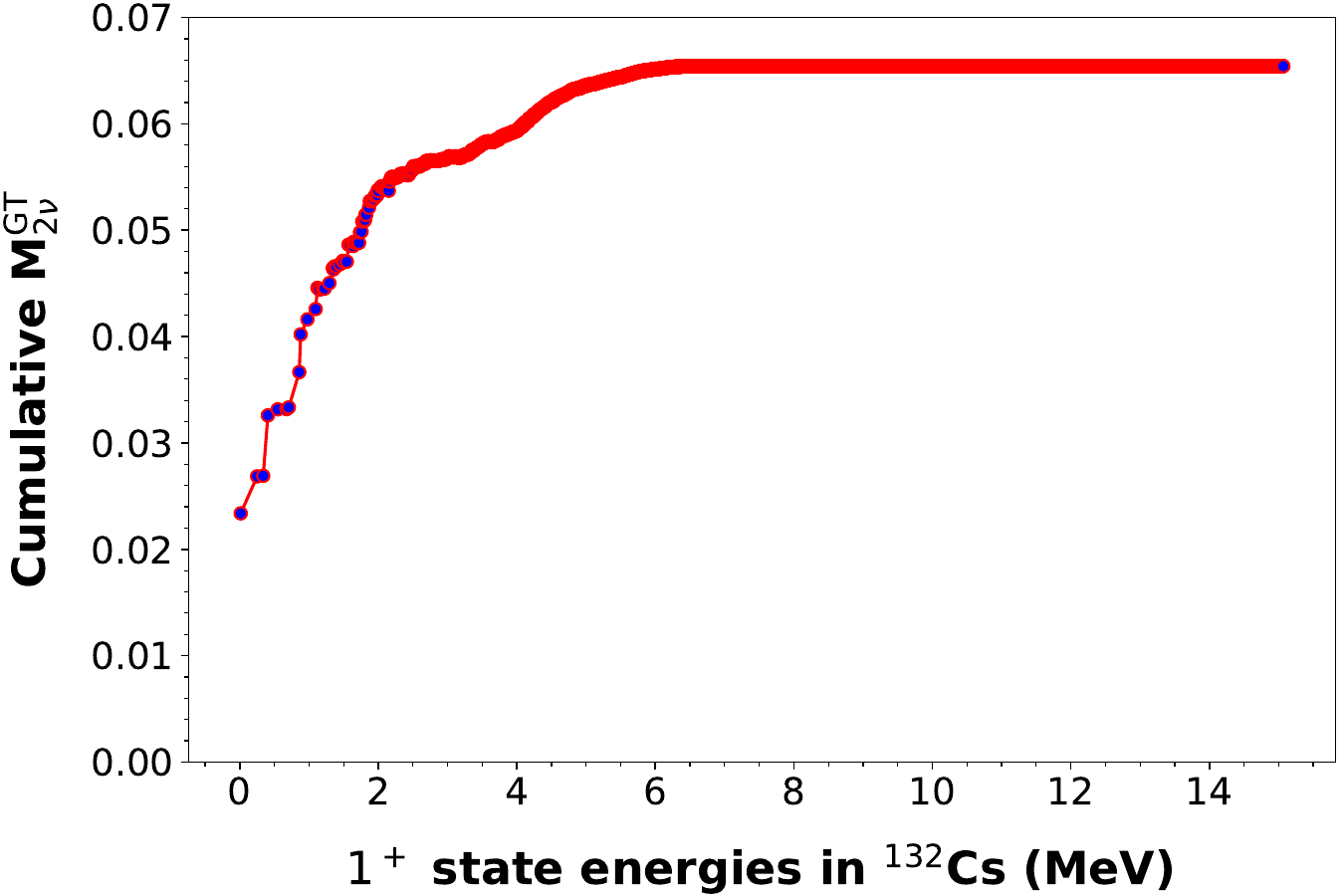}
	\caption{\label{NME_variation_132Ba} Variation of the cumulative NME for the GT transitions with respect to the $1^+$ state energies in the intermediate nucleus $^{132}$Cs.}
\end{figure}

Here, we analyze the variation of the cumulative NMEs for the GT transitions associated with the $2\nu$ECEC process in $^{132}$Ba as a function of the $1^+$ state energies in the intermediate nucleus $^{132}$Cs, as shown in Fig. \ref{NME_variation_132Ba}, along with the corresponding half-life ($T_{1/2}^{2\nu}$) calculated using the final NME (see Table \ref{table:half_life_132Ba}). The shell-model calculations successfully reproduce the $2^+_{\rm g.s.}$ level in $^{132}$Cs, and predict the $1^+_1$ state at 13 keV above the $2^+_{\rm g.s.}$. Since no experimental $1^+$ states in $^{132}$Cs have been confirmed to date, we use shell-model predicted energies in our calculations. The lowest $1^+$ state contributes 35.8\% to the final NME (0.0654) with a value of 0.0234. With increasing $1^+$ state energies, the cumulative NME increases for most of the energies and saturates around $\sim6.3$ MeV with a final value of 0.0654 (see Fig. \ref{NME_variation_132Ba}). To test the reliability of the shell-model predicted $1^+_1$ state energy in $^{132}$Cs, we varied its location from 13 to 513 keV (a range of 500 keV) in steps of 50 keV and accordingly shifted the energies of the other $1^+$ states relative to it. We found that as the $1^+_1$ state energy increases up to 100 keV (from 13 to 113 keV), the final NME changes only slightly (from 0.0654 to 0.0628), whereas at 513 keV, a significant suppression appears, reducing the NME to 0.0542. This indicates that if the $1^+_1$ state is experimentally confirmed below 100 keV in the future, the accuracy of the present NME calculation will be maintained. However, if it lies above this range, the NME would gradually be affected, becoming significantly reduced near or beyond 500 keV. Nevertheless, such a situation should be unlikely, as the other shell-model predicted states show reasonable agreement with the available experimental data, although precise measurement of the $1^+_1$ state energy in $^{132}$Cs still remains desirable.

\begin{table} 
	\centering
	\caption{Shell-model calculated $2\nu$ECEC NME and the extracted half-life for $^{132}$Ba.}
	\begin{tabular}{ccccc}
		\hline \hline
		$|\rm NME|$ & \makecell{$G_{2\nu}^{\rm ECEC}$ \\ (yr$^{-1}$) \cite{Nitescu}} & $g_A^{\rm eff}$ & \makecell{Calculated \\ $T_{1/2}^{2\nu}$ (yr)} & \makecell{Experimental \\ $T_{1/2}^{2\nu}$ (yr) \cite{Meshik}} \\
		\hline
		
		0.0654   & \makecell{$40.87$ \\ $\times 10^{-24}$}   & 0.94 & 7.33$\times 10^{24}$ & $> 2.2\times 10^{21}$  \\
		
		& & 0.72 & 2.13$\times 10^{25}$ & \\
		
		& & 0.57 & 5.42$\times 10^{25}$ & \\
		\hline \hline 
	\end{tabular}
	\label{table:half_life_132Ba}
\end{table}

\begin{figure}
	\centering
	\includegraphics[width=70mm]{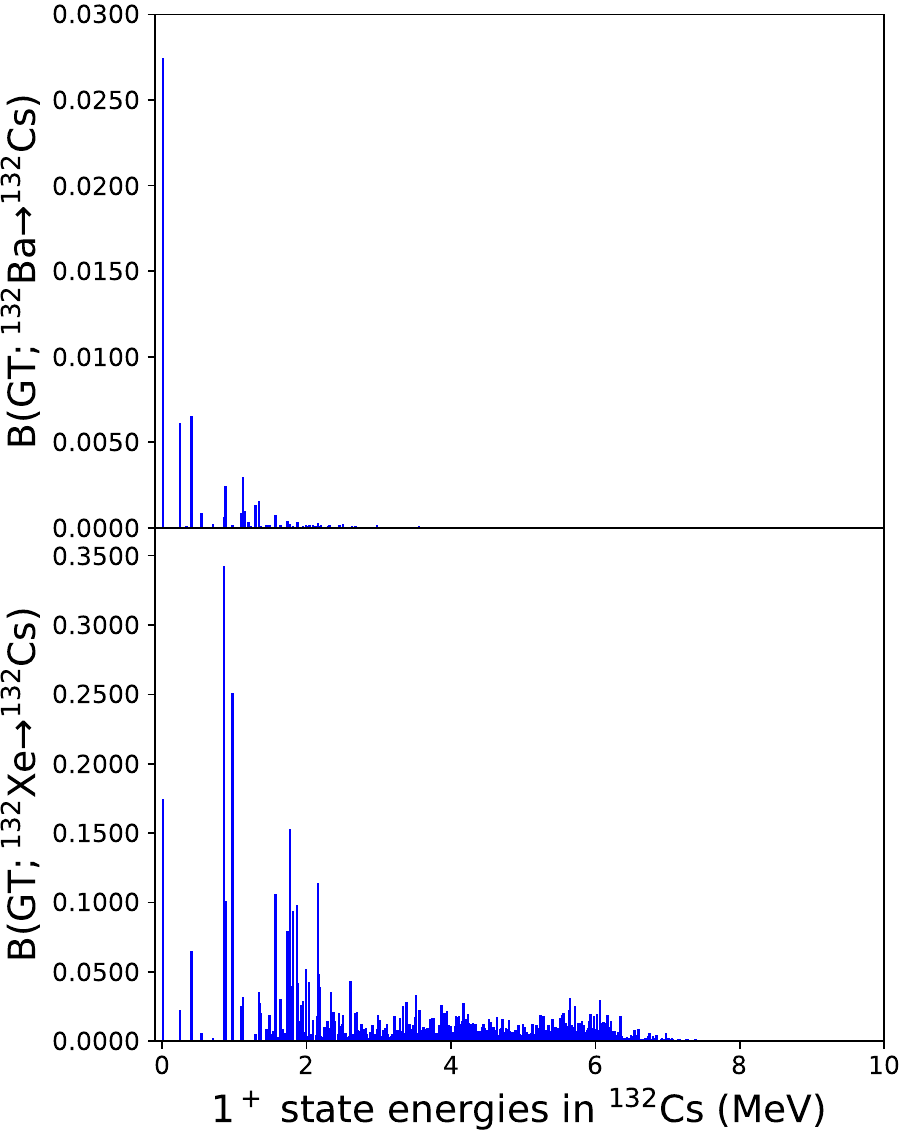}
	\caption{\label{B_GT_132Ba} Calculated GT transition strengths $B(GT)$ for the $^{132}$Ba$(0^+_{\rm g.s.})$ $\rightarrow$ $^{132}$Cs$(1^+_{n})$ (upper panel) and $^{132}$Xe$(0^+_{\rm g.s.})$ $\rightarrow$ $^{132}$Cs$(1^+_{n})$ (lower panel) transitions as a function of the excitation energy of the intermediate $1^+$ states in $^{132}$Cs.}
\end{figure}

For the mass range $A=130$-$136$, in the proton and neutron valence shells $0g_{7/2}1d2s0h_{11/2}$, the value of $g_{\rm A}^{\rm eff}$ was reported as 0.94 in Ref. \cite{Suhonen4}, based on a phenomenological quenching of the GT operator ($q=0.74$) inferred from shell-model studies of $\beta\beta$ decay in nuclei such as $^{124}$Sn \cite{Horoi}, $^{130}$Te, and $^{136}$Xe \cite{Neacsu}. We have also plotted the calculated GT strength distribution in Fig. \ref{B_GT_132Ba} with $g_{\rm A}^{\rm eff}=0.94$ for the $^{132}$Ba$(0^+_{\rm g.s.})$ $\rightarrow$ $^{132}$Cs$(1^+_{n})$ and $^{132}$Xe$(0^+_{\rm g.s.})$ $\rightarrow$ $^{132}$Cs$(1^+_{n})$ transitions. Thus, we have reported the half-life of $^{132}$Ba for $2\nu$ECEC in Table \ref{table:half_life_132Ba} with a value of $g_{\rm A}^{\rm eff}=0.94$, using the final NME with GT transitions, $M_{2\nu}^{\rm GT}=0.0654$ (calculated from Eq. (\ref{eq2})), and $G_{2\nu}^{\rm ECEC}=40.87\times 10^{-24}$ yr$^{-1}$ \cite{Nitescu}. We found that our estimated value of $T_{1/2}^{2\nu}$ is $7.33\times 10^{24}$ yr, which exceeds the currently available experimental lower limit of $2.2\times 10^{21}$ yr \cite{Meshik} by approximately three orders of magnitude. We have also computed the half-life for $g_{\rm A}^{\rm eff}=0.72$ and 0.57, values adopted in our previous study \cite{Patel_NPA} for the same mass region, obtaining $T_{1/2}^{2\nu}=2.13\times 10^{25}$ and $5.42\times 10^{25}$ yr, respectively.

Over the past few decades, experimental investigations for the half-life measurement of the $2\nu$ECEC process in $^{132}$Ba have been very limited. Therefore, future measurements with greater precision may refine the current estimate ($> 2.2\times 10^{21}$ yr) and establish an improved, possibly higher lower bound for the half-life. However, theoretical calculations in a more enlarged model space, though very challenging at present, may further improve the NME and half-life predictions in the future, once substantially more advanced computational resources become available.

\begin{figure*}
	\centering
	\includegraphics[width=40mm]{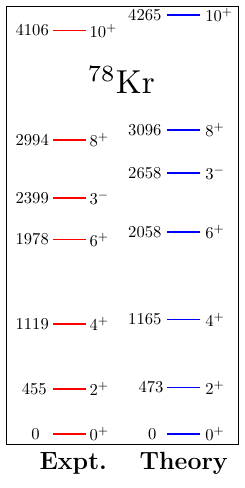}
	\includegraphics[width=40mm]{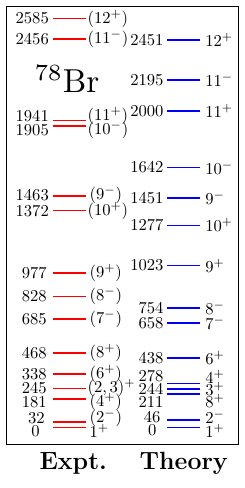}
	\includegraphics[width=40mm]{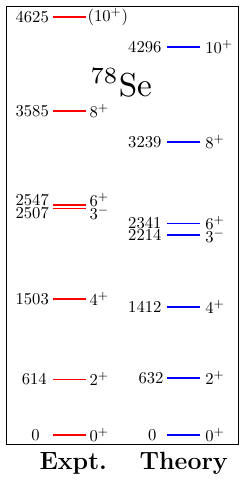}
	\caption{\label{spectrum_78Kr} Comparison between the theoretical and experimental \cite{NNDC} energy levels (in keV) in the $^{78}$Kr, $^{78}$Br and $^{78}$Se. The calculated energies of $^{78}$Br are shown with respect to the lowest $1^+$ state.}
\end{figure*}

\subsection{Improved results for the spectroscopic properties and $2\nu$ECEC in $^{78}$Kr} \label{Sec-IIIC}

\begin{table*}
	\centering
	\caption{Comparison between theoretical and experimental $B(E2)$ strengths [in Weisskopf units (W.u.)] for $^{78}$Kr, $^{78}$Br and $^{78}$Se, together with the comparison of theoretical and experimental $Q_s$ values [in $e$b] for $^{78}$Se. The used parameters are taken to be the same as in Table \ref{table:BE2}. Additionally, the shell-model predicted $B(E2)$ values from our previous study \cite{Patel_78Kr} are listed in the last column for comparison.}
	\begin{tabular}{cccccc}
		\hline \hline
		
		&  &  & $B(E2)$ & &  \\
		
		\hline
		Nucleus & $J^{\pi}_i \to J^{\pi}_f$ & \multicolumn{2}{c}{Theory} & Expt. \cite{NNDC_NUDAT} & Previous results \cite{Patel_78Kr} \\
		\hline 
		
		&  & $(1.6, 0.8)e$ & $(1.7, 1.1)e$ &  &  $(1.7, 1.1)e$ \\
		
		\cline{3-4}
		\cline{6-6}
		
		$^{78}$Kr   & $2^+_1 \to 0^+_1$   & 35.2 & 45.4 & 67.9(22) & 43.3 \\
		& $4^+_1 \to 2^+_1$   & 50.7 & 65.4 & 88(5) & 62.5 \\
		& $6^+_1 \to 4^+_1$   & 55.4 & 71.6 & 94(11) & 68.2 \\
		& $8^+_1 \to 6^+_1$   & 54.2 & 70.4 & $\approx$85 & 67.5 \\
		& $10^+_1 \to 8^+_1$   & 52.9 & 68.8 & 80(12) & 63.8 \\
		\hline \\
		
		$^{78}$Br   & $6^+_1 \to 4^+_1$   & 15.4 & 21.0 & - & - \\
		
		& $8^-_1 \to 7^-_1$  & 14.1 & 18.6 & - & - \\
		
		& $9^-_1 \to 7^-_1$  & 32.7 & 42.6 & - & - \\
		
		& $10^+_1 \to 8^+_1$  & 20.7 & 27.7 & - & - \\
		
		& $10^-_1 \to 8^-_1$  & 57.1 & 74.6 & - & - \\
		
		\hline \\
		
		$^{78}$Se   & $2^+_1 \to 0^+_1$   & 25.6 & 35.0 & 33.5(8) & 28.3 \\
		& $4^+_1 \to 2^+_1$   & 35.3 & 48.3 & 49.5(24) & 38.6 \\
		& $6^+_1 \to 4^+_1$   & 36.1 & 49.0 & 47(14) & 38.4 \\
		& $8^+_1 \to 6^+_1$   & 37.8 & 50.6 &  56(19) & 33.9 \\
		& $10^+_1 \to 8^+_1$   & 47.1 & 61.9 & - & 42.7 \\
		\hline
		
		&  &  & $Q_s$ & &  \\
		
		\hline
		
		Nucleus & $J^{\pi}$ & \multicolumn{2}{c}{Theory} & Expt. \cite{Hayakawa} & - \\
		
		\hline
		
		&  & $(1.6, 0.8)e$ & $(1.7, 1.1)e$ &  &  - \\
		
		\cline{3-4}
		
		$^{78}$Se & $2^+_1$ & -0.329 & -0.387 & -0.20$\pm$0.07 & - \\
		
		& $4^+_1$ & -0.391 & -0.462 & -0.68$\pm$0.15 & - \\
		
		& $6^+_1$ & -0.153 & -0.186 & - & - \\
		
		\hline \hline
		
	\end{tabular}
	\label{table:BE2_78Kr}
\end{table*}

In Fig. \ref{spectrum_78Kr}, we have presented a comparison between the theoretical and experimental low-lying energy levels of the parent ($^{78}$Kr), intermediate ($^{78}$Br), and granddaughter ($^{78}$Se) nuclei for $2\nu$ECEC. The present shell-model calculations exhibit quite remarkable agreement with the experimental data for these nuclei. In our previous work \cite{Patel_78Kr}, the calculated $8^+_1$ and $10^+_1$ states in $^{78}$Kr were primarily dominated by the $\pi(f_{5/2}^3p_{3/2}^3g_{9/2}^2) \otimes \nu(g_{9/2}^4)$ configuration, which differed from the structure of the lower-lying $0^+_1$, $2^+_1$, $4^+_1$, and $6^+_1$ states. These lower states ($0^+_1$, $2^+_1$, $4^+_1$, and $6^+_1$) were mainly characterized by the $\pi(f_{5/2}^4p_{3/2}^2g_{9/2}^2) \otimes \nu(g_{9/2}^4)$ and $\pi(f_{5/2}^2p_{3/2}^4g_{9/2}^2) \otimes \nu(g_{9/2}^4)$ configurations, which are consistent with the corresponding states in the present study. In the current analysis, the $8^+_1$ and $10^+_1$ states are also predominantly governed by the $\pi(f_{5/2}^4p_{3/2}^2g_{9/2}^2) \otimes \nu(g_{9/2}^4)$ configuration, supporting the fact that the $0^+_1$-$10^+_1$ states in $^{78}$Kr belong to the same band. The present calculations yield positive $Q_s$ values for these $2_1^+$, $4_1^+$, $6_1^+$, $8_1^+$, and $10_1^+$ states and support negative $\beta_2=-0.27$, $-0.26$, $-0.23$, $-0.17$, and $-0.11$, respectively (for $K=0$) with the appropriate effective charges $(e_\pi,e_\nu)=(1.7,1.1)e$, which indicates an oblate intrinsic shape. The gradual reduction of $|\beta_2|$ with increasing spin further suggests possible shape mixing. Our results are consistent with the study of Sun $et$ $al.$ \cite{Sun}, but contrast with the predominantly prolate character inferred by Becker $et$ $al.$ \cite{Becker}. Although neutron-deficient Kr isotopes exhibit well-known shape coexistence, different studies have suggested competing prolate and oblate configurations \cite{Becker}. For the case of $^{78}$Se, the dominant configurations of the $0^+_1$-$10^+_1$ states are quite similar to those predicted in the previous calculations. In our calculations for $^{78}$Se, the predicted $Q_s$ values for the $2^+_1$, $4^+_1$, and $6^+_1$ states are negative, consistent with the observed $Q_s$ values available for the $2^+_1$ and $4^+_1$ states \cite{Hayakawa}. The comparison between the shell-model predicted and experimental $Q_s$ values for these states is reported in Table \ref{table:BE2_78Kr}. However, the predicted $Q_s$ values for the $8^+_1$ and $10^+_1$ states are positive. Therefore, we have also calculated the $8^+_2$ and $10^+_2$ states, located at 3858 and 4875 keV, respectively, which are closer to the observed $8^+$ and $10^+$ states of the g.s. band than the predicted $8^+_1$ and $10^+_1$ states. The $Q_s$ values for the $8^+_2$ and $10^+_2$ states are negative, similar to those of the $2^+_1$, $4^+_1$, and $6^+_1$ states. The negative $Q_s$ values for the $2^+_1$, $4^+_1$, $6^+_1$, $8^+_2$, and $10^+_2$ states imply positive $\beta_2$ values, indicating prolate deformation, consistent with $\beta_2$ value of the previous study \cite{Moreno}. In the odd-$Z$ and odd-$N$ intermediate nucleus $^{78}_{35}$Br$_{43}$, the shell-model predicted low-lying energy levels are plotted relative to the $1^+_1$ state for comparison with the experimental data (see Fig. \ref{spectrum_78Kr}). Although the calculated g.s. is $2^-$, the $1^+_1$ state lies 0.372 MeV higher in energy. Most of the calculated excited states show quite reasonable agreement with the experimental levels, which strengthens the predictive efficiency of the employed GWBXG effective interaction. Previously, we compared the theoretical and experimental $B(E2)$ strengths in $^{78}$Kr and $^{78}$Se using effective charges $(1.7, 1.1)e$. In the present study, the calculated $B(E2)$ values show further improvement over the earlier results \cite{Patel_78Kr}, achieving better consistency with the experimental data, as reported in Table \ref{table:BE2_78Kr}, while the excitation energies do  not show much improvement, as the earlier calculations already reproduced the experimental spectra well. This improved agreement supports the adequacy of the present model space for describing the underlying nuclear structure relevant to the $2\nu$ECEC process in $^{78}$Kr. Similar to $^{132}$Cs, the experimental $B(E2)$ values are unavailable for any states in $^{78}$Br; therefore, only the shell-model predictions are reported.

\begin{figure}
	\centering
	\includegraphics[width=86mm]{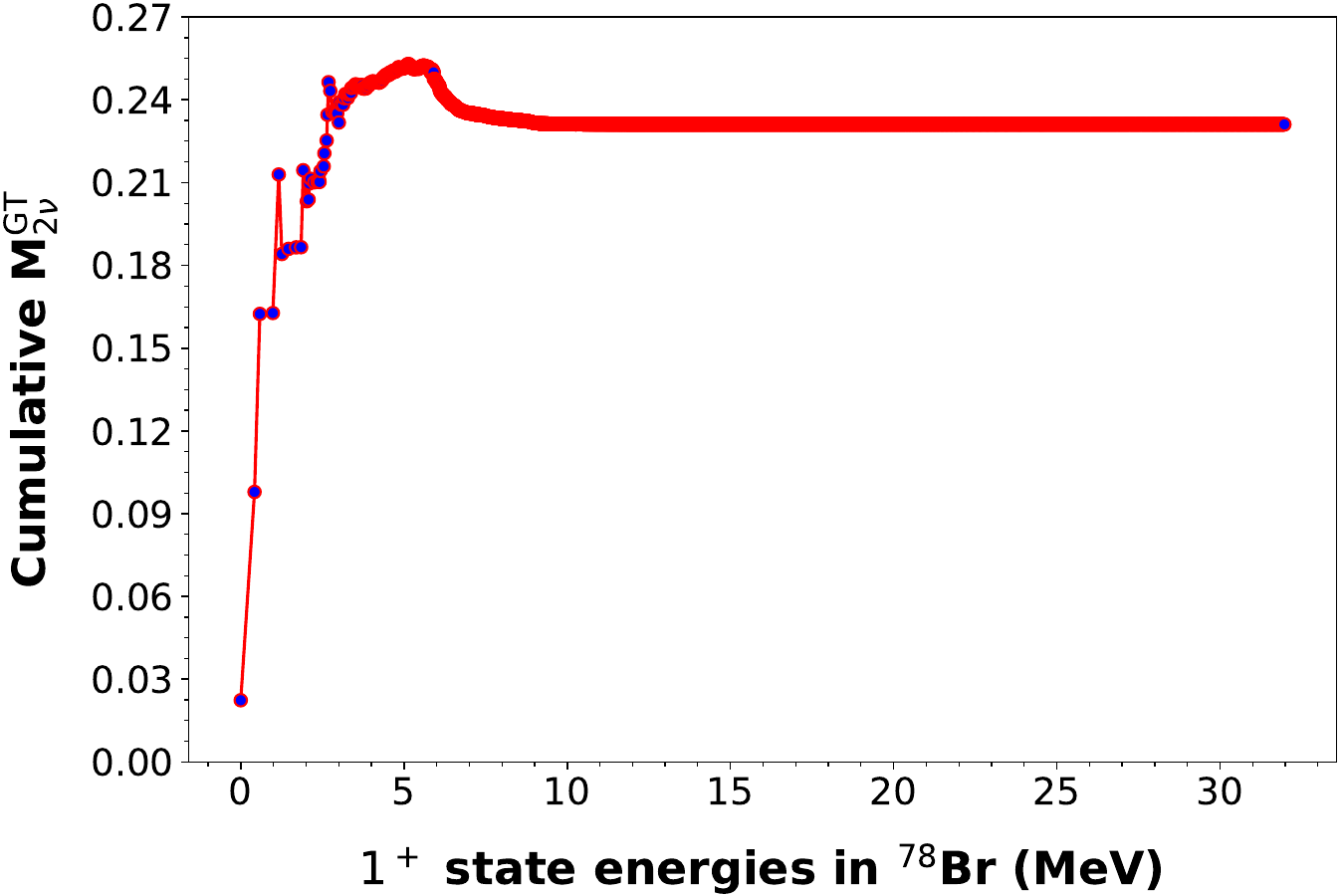}
	\caption{\label{NME_variation_78Kr} Variation of the cumulative NME for the GT transitions with respect to the $1^+$ state energies in the intermediate nucleus $^{78}$Br.}
\end{figure}

\begin{table*} 
	\centering
	\caption{Shell-model calculated $2\nu$ECEC NME and the extracted half-life for $^{78}$Kr.}
	\begin{tabular}{ccccc}
		\hline \hline
		$|\rm NME|$ & $G_{2\nu}^{\rm ECEC}$ (yr$^{-1}$) \cite{Nitescu} & $g_A^{\rm eff}$ & Calculated $T_{1/2}^{2\nu}$ (yr) & Experimental $T_{1/2}^{2\nu}$ (yr) \\
		\hline
		
		0.2311   & $520.518 \times 10^{-24}$   & 0.8 & 8.78$\times 10^{22}$ & \makecell{$[9.2^{+5.5}_{-2.6}({\rm stat}) \pm  1.3({\rm syst})]$  $\times 10^{21}$ \cite{Gavrilyuk}} \\
		
		& & & & \makecell{$[1.9^{+1.3}_{-0.7}({\rm stat}) \pm 0.3 ({\rm syst})]$  $\times 10^{22}$ \cite{Ratkevich}} \\
		
		\hline \hline
	\end{tabular}
	\label{table:half_life_78Kr}
\end{table*}

As depicted in the  Fig. \ref{NME_variation_78Kr}, the NME associated with GT transitions, calculated using Eq. (\ref{eq2}), is plotted as a function of the $1^+$ state energies in $^{78}$Br. In our calculations, the excitation energies of the $1^+$ states in $^{78}$Br are adjusted to place the lowest $1^+$ state at the experimental energy of 0 MeV. The contributions from these $1^+$ states are predominantly constructive up to approximately 5.15 MeV, where the cumulative NME reaches its maximum value of 0.2529. Beyond this energy, the cumulative NME exhibits mostly a decline up to about 11 MeV; after that, it begins to saturate with a final value of 0.2311. Notably, the pronounced peak observed around 5.15 MeV in the present work was absent in our previous study \cite{Patel_78Kr}. Moreover, the final NME value of 0.2311 is enhanced relative to the previously reported value of 0.1997 \cite{Patel_78Kr}, and it is noteworthy that the NME extracted from the recently measured half-life \cite{Ratkevich,Nitescu}, $0.318^{+0.100}_{-0.073}$, exceeds our predicted value by only a factor of $\sim$1.38. This improvement is most likely due to the use of a larger model space in the present calculations.

\begin{figure}
	\centering
	\includegraphics[width=70mm]{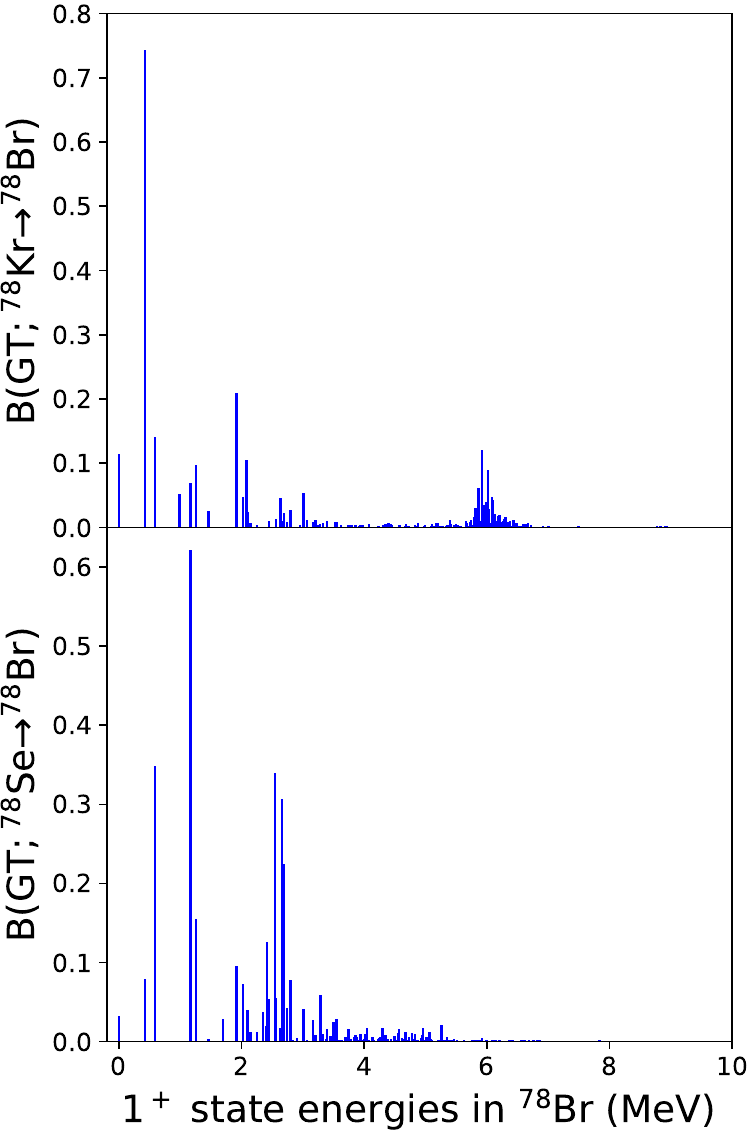}
	\caption{\label{B_GT_78Kr} Calculated GT transition strengths $B(GT)$ for the $^{78}$Kr$(0^+_{\rm g.s.})$ $\rightarrow$ $^{78}$Br$(1^+_{n})$ (upper panel) and $^{78}$Se$(0^+_{\rm g.s.})$ $\rightarrow$ $^{78}$Br$(1^+_{n})$ (lower panel) transitions as a function of the excitation energy of the intermediate $1^+$ states in $^{78}$Br.}
\end{figure}

For the wide mass range $A=63-96$, $g_{\rm A}^{\rm eff} = 0.8$ was adopted, as reported in Ref. \cite{Suhonen4} and derived from the earlier work \cite{Honma_ga}. This effective value corresponds to a phenomenological quenching of the GT operator determined from comparisons between shell-model calculations and experimental data for this mass region and applied to calculate $2\nu\beta^-\beta^-$ observables in nuclei such as $^{76}$Ge and $^{82}$Se \cite{Honma_ga}. We have also shown calculated GT strength distribution in Fig. \ref{B_GT_78Kr} corresponding to $g_{\rm A}^{\rm eff} = 0.8$ for the $^{78}$Kr$(0^+_{\rm g.s.})$ $\rightarrow$ $^{78}$Br$(1^+_{n})$ and $^{78}$Se$(0^+_{\rm g.s.})$ $\rightarrow$ $^{78}$Br$(1^+_{n})$ transitions. Employing $g_{\rm A}^{\rm eff}$ value as 0.8, along with the calculated NME corresponding to the GT transitions (0.2311) and the $G_{2\nu}^{\rm ECEC}$ taken from Ref. \cite{Nitescu}, the extracted half-life was found as $8.78 \times 10^{22}$ yr. Here, we have used the $G_{2\nu}^{\rm ECEC}$ value corresponding to the $2\nu$KK mode, as the experimental half-lives reported in Table \ref{table:half_life_78Kr} are related to electron captures from K-shell pairs \cite{Gavrilyuk,Ratkevich}. Employing the same $G_{2\nu}^{\rm ECEC}$ value and $g_{\rm A}^{\rm eff}=0.8$ along with our previously obtained NME (0.1997) \cite{Patel_78Kr}, we calculated a half-life as $11.76 \times 10^{22}$ yr. This value is less close to the recently measured half-life, $[1.9^{+1.3}_{-0.7}({\rm stat}) \pm 0.3({\rm syst})] \times 10^{22}$ yr \cite{Ratkevich}, compared to the present result (see Table \ref{table:half_life_78Kr}).

In addition to the $2\nu$ECEC mode, the decay of $^{78}$Kr can also proceed through competing positron-emitting channels, namely the $2\nu\beta^+\beta^+$ and the $2\nu{\rm EC}\beta^+$ decays. While these decay modes share the same NMEs from our calculations, they differ in their corresponding phase-space factors. Using the calculated GT NME (0.2311) and $g_{\rm A}^{\rm eff}=0.8$, we estimate the half-lives for the $2\nu\beta^+\beta^+$ and $2\nu{\rm EC}\beta^+$ decay modes of $^{78}$Kr to be $4.68\times10^{26}$ and $1.19\times10^{23}$ yr, respectively, employing the $G_{2\nu}^{\beta^+\beta^+}$ and $G_{2\nu}^{{\rm EC}\beta^+}$ from Ref. \cite{Kotila}. The experimental lower limits of the half-lives are $2.0\times10^{21}$ and $1.1\times10^{20}$ yr, corresponding to the $2\nu\beta^+\beta^+$ and $2\nu{\rm K}\beta^+$ channels, respectively \cite{Saenz}.

It should be noted that the present shell-model calculations may contain small isospin-symmetry breaking effects arising from the employed effective interactions and adopted model spaces. In the $^{132}$Ba case, the SN100PN interaction contains an explicit Coulomb term in the proton-proton TBMEs, and the proton and neutron model spaces are not exactly identical ($\nu g_{7/2}$ orbital is completely occupied in the calculations). Similarly, for $^{78}$Kr, the adopted GWBXG interaction contains non-identical proton-proton and neutron-neutron TBMEs in the shared $p_{1/2}$ and $g_{9/2}$ orbitals, with Coulomb effects effectively embedded in the parametrization, while the proton and neutron valence spaces are also different ($Z=28-50$ and $N=38-70$). These effects may introduce some uncertainty in the calculated GT NMEs.

\section{Summary and Conclusions} \label{section4}

In this work, we have performed large-scale shell-model calculations to investigate the $2\nu$ECEC process in $^{132}$Ba and $^{78}$Kr using the SN100PN and GWBXG effective interactions, respectively. The model space employed with the SN100PN interaction consists of the $0g_{7/2}1d2s0h_{11/2}$ proton and neutron orbitals. While, for the GWBXG interaction, the model space includes the $0f_{5/2}1p0g_{9/2}$ proton orbitals and the $1p_{1/2}0g1d2s$ neutron orbitals. As a necessary consistency check, the spectroscopic properties of the parent, intermediate, and granddaughter nuclei involved in the decay were examined, showing good agreement with the experimental excitation energies and $B(E2)$ transition strengths, thereby supporting the reliability of the underlying shell-model wave functions. Based on these wave functions, the NMEs and corresponding half-lives were evaluated. The convergence behavior of the cumulative GT matrix elements was further analyzed as a function of the excitation energies of the intermediate $1^+$ states. The resulting NME for $^{132}$Ba provides a baseline theoretical estimate that can assist future experimental efforts in constraining improved half-life limits; its robustness was further examined by exploring the sensitivity to the shell-model energy of the unconfirmed $1_1^+$ state in $^{132}$Cs. The predicted half-life for the $2\nu$ECEC process in $^{132}$Ba obtained using GT transitions is $7.33\times10^{24}$ yr (with $g_{\rm A}^{\rm eff}=0.94$), which is consistent with the presently known experimental lower limit of $2.2\times10^{21}$ yr. For $^{78}$Kr, the present results represent an updated and improved shell-model estimate relative to earlier calculation within more restricted model space. The predicted half-life for the $2\nu$ECEC process in $^{78}$Kr from GT transitions is $8.78\times10^{22}$ yr (with $g_{\rm A}^{\rm eff}=0.8$), in reasonable agreement with the experimentally observed value of $[1.9^{+1.3}_{-0.7}({\rm stat}) \pm 0.3 ({\rm syst})]\times10^{22}$ yr. The present study employs the standard one-body GT operator and therefore does not explicitly incorporate all higher-order many-body effects. Nevertheless, the use of effective axial couplings in the half-life calculations allows the present predictions to be regarded as reasonable within the typical theoretical uncertainties associated with the $2\nu$ECEC process.

In future work, it would be desirable to perform calculations in larger model spaces including almost all relevant spin-orbit partner orbitals for both protons and neutrons. In addition, effective interactions with improved isospin symmetry could help constrain possible isospin-symmetry breaking effects in the calculated NMEs.

\section*{Acknowledgements}

We acknowledge financial support from the Science and Engineering Research Board (SERB), India, through the research grant CRG/2022/005167. We would also like to thank the National Supercomputing Mission (NSM) for providing computing resources of ``PARAM Ganga'' at the IIT Roorkee, implemented by C-DAC and supported by the Ministry of Electronics and Information Technology (MeitY) and the Department of Science and Technology (DST), Government of India.

\end{document}